\documentclass[a4paper,11pt]{article}
\pdfoutput=1 % if your are submitting a pdflatex (i.e. if you have images in pdf, png or jpg format)

\usepackage{jinstpub} % for details on the use of the package, please see the JINST-author-manual

\usepackage{hyperref}
\usepackage{lineno}
\usepackage{url}

\title{\boldmath Effects of lightning strokes on underground gravitational waves observatories}

\author[a,1]{T. Washimi\note{Corresponding author.}}
\author[b]{T. Yokozawa,}
\author[b]{M. Nakano,}
\author[b]{T. Tanaka,}
\author[c]{K. Kaihotsu,}
\author[c]{Y. Mori,}
\author[d]{and T. Narita}

\affiliation[a]{Gravitational Wave Science Project (GWSP), Kamioka branch, National Astronomical Observatory of Japan (NAOJ),\\ Kamioka-cho, Hida City, Gifu 506-1205, Japan}
\affiliation[b]{Institute for Cosmic Ray Research (ICRR), KAGRA Observatory, the University of Tokyo,\\ Kamioka-cho, Hida City, Gifu 506-1205, Japan}
\affiliation[c]{Department of Physics, the University of Toyama,\\ Gofuku, Toyma City, Toyama 930-8555, Japan}
\affiliation[d]{Department of Electrical and Electric Engineering, Shonan Institute of Technology,\\ Tsujido, Fujisawa City, Kanagawa 251-8511, Japan}

\emailAdd{tatsuki.washimi@nao.ac.jp}

\abstract{
For ground-based gravitational wave (GW) detectors, lightning strokes in the atmosphere are sources of environmental noise. 
Some GW detectors are built or planned in underground facilities, and knowledge of how lightning strokes affect them is of interest.
In this paper, the lightning detection system in KAGRA is introduced, and the properties of the magnetic field measured inside and outside the KAGRA tunnel are shown. 
One lightning-induced event in the GW channel of the KAGRA main interferometer is also  showed. 
Finally, possible applications of lightning events for the GW experiments are discussed.
}

\keywords{%Only keywords from JINST's keywords list please : https://jinst.sissa.it/jinst/help/keywordsList.jsp
Large detector systems for particle and astroparticle physics, 
Interferometry, 
Instrumental noise
}

\arxivnumber{2103.06516} % only if you have one

\begin{document}
\maketitle
\flushbottom

%%%%%%%%%%%%%%%%%%%%%%%%%%%%%%%%%%%%%%%%%%%%%%%%%%%%%%%%%%%%%%%%%%%%%%%%%%%%%%%%%%%%%%%%%%
\section{Gravitational wave and underground facilities}
\label{sec:gw-underground}

\begin{figure}[htbp]
\centering 
\includegraphics[width=15cm,clip]{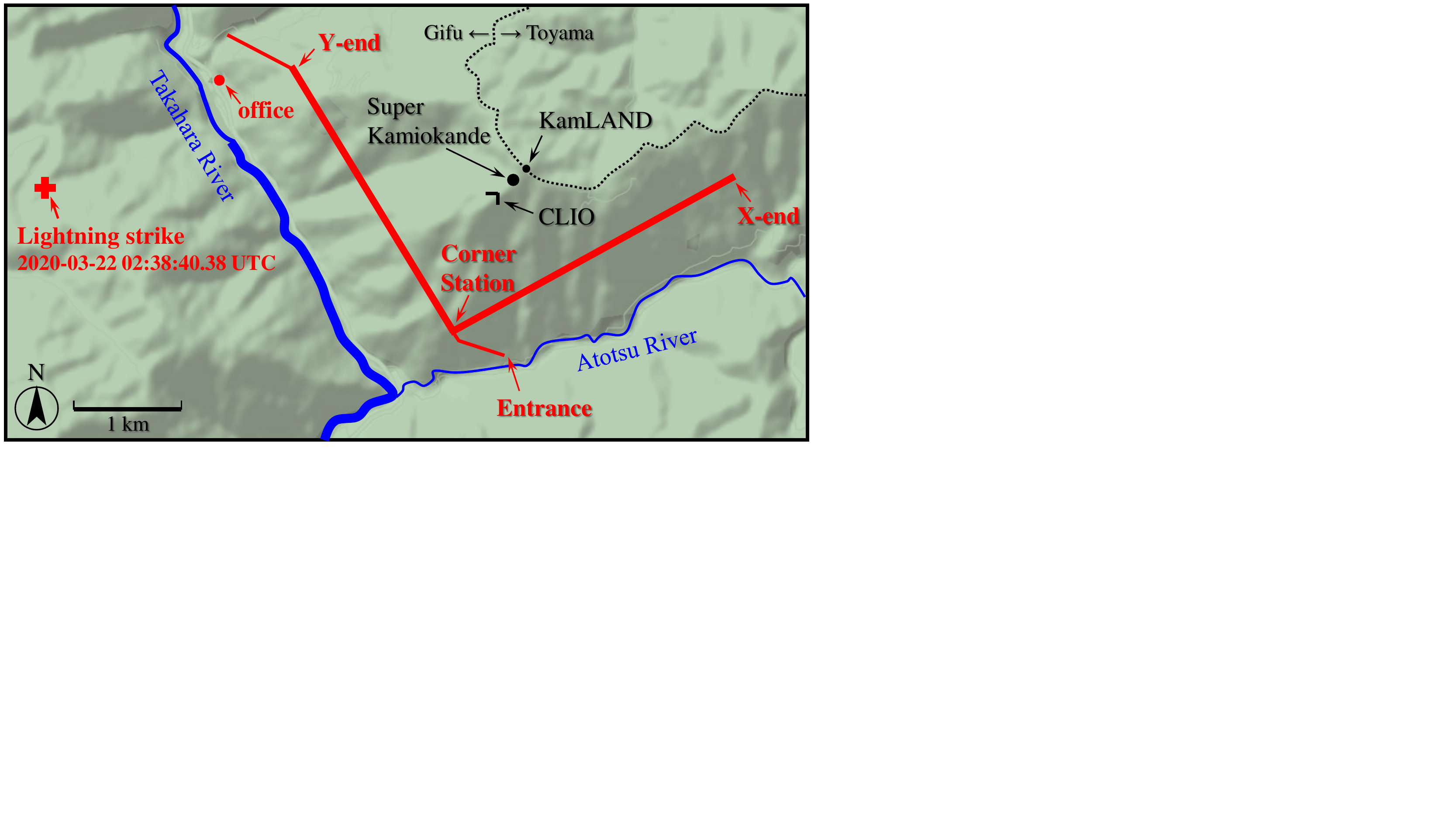}
\caption{\label{fig:kamioka} The map of the KAGRA experimental site and nearby area. 
The lightning stroke on 2020-03-22 02:38:40.38 UTC is discussed in Section~\ref{sec:3/22event}.}
\end{figure}

To detect a Gravitational Wave (GW) with high signal-to-noise ratio, locating the interferometric detector underground is an effective way to avoid and filtering noise sources caused by human activities and meteorological phenomenon. 
For example, seismic noise can be reduced by approximately $\mathcal{O}(10^2)$ for underground facilities compared to the surface~\cite{Tomaru}.   
KAGRA~\cite{PTEP01} and its prototype CLIO~\cite{CLIO} have been built in the underground facilities of Kamioka (Figure~\ref{fig:kamioka}), Japan. 
KAGRA conducted the first joint observation run (O3GK) with GEO600 in Germany from April 7 to 21, 2020~\cite{O3GK}. 
In the next future the third generations detectors for Gravitational Waves continuous studies and observations, like ET (Einstein Telescope)~\cite{ET}  in Europe and ZAIGA~\cite{ZAIGA} in China, are planned to be accommodated within giant laboratories underground.

Because the GW signals are very small, having a dimensionless amplitude $h$ of the order of  $10^{-21}$-$10^{-22}$, it is extremely important to have an environmental monitoring system studying human activities, local disturbances and geophysical phenomena.
Physical environmental monitoring (PEM)\footnote{Details of the KAGRA PEM have been reported in a review paper~\cite{PTEP3}.} 
is one part of the GW observation conducted to monitor such noise with seismometers, accelerometers, microphones, and magnetometers. 
In this study, we focus on the magnetic field generated by lightning strokes. 
Lightning generates a large pulse of a magnetic field and possibly affects the GW signal by shaking the mirror, or incurs electrical noise. 
However, its effect underground is non-trivial. 
In the KAGRA experimental site, there are three 3-axial fluxgate magnetometers (Bartington Mag-13 MCL100\textregistered, with a frequency range of DC-3~kHz) 
located in the corner station (CS), in the X-end, and in the Y-end (Figure~\ref{fig:kamioka}), respectively. 
In addition, a lightning detector is located in front of the KAGRA tunnel, and its details are described in the next section.

%%%%%%%%%%%%%%%%%%%%%%%%%%%%%%%%%%%%%%%%%%%%%%%%%%%%%%%%%%%%%%%%%%%%%%%%%%%%%%%%%%%%%%%%%%
%\newpage
\section{Localization of the lightning strokes by Blitzortung network}
\label{sec:blitz}

\begin{figure}[htbp]
\centering 
\includegraphics[width=6cm,clip]{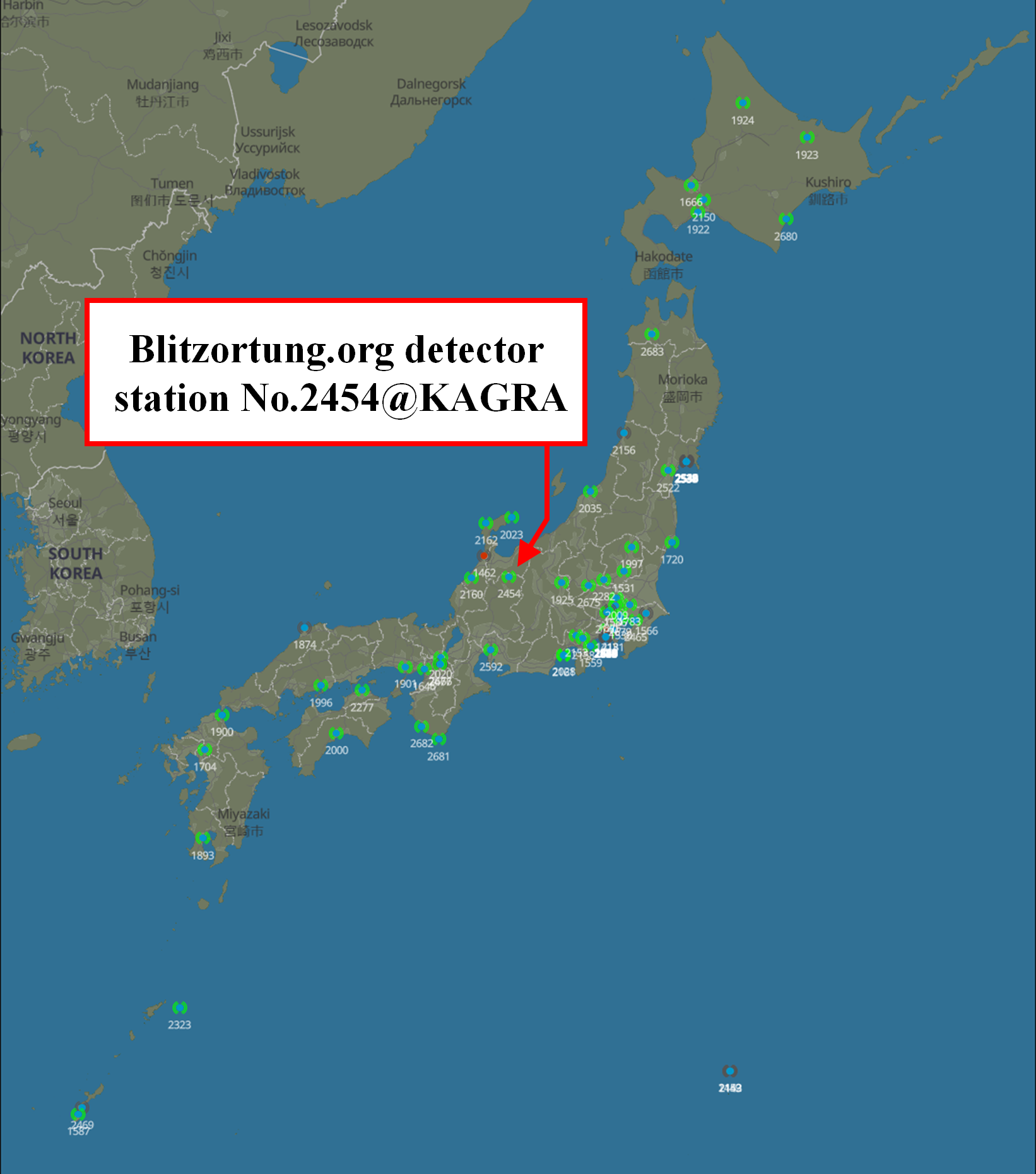}
\includegraphics[width=8cm,clip]{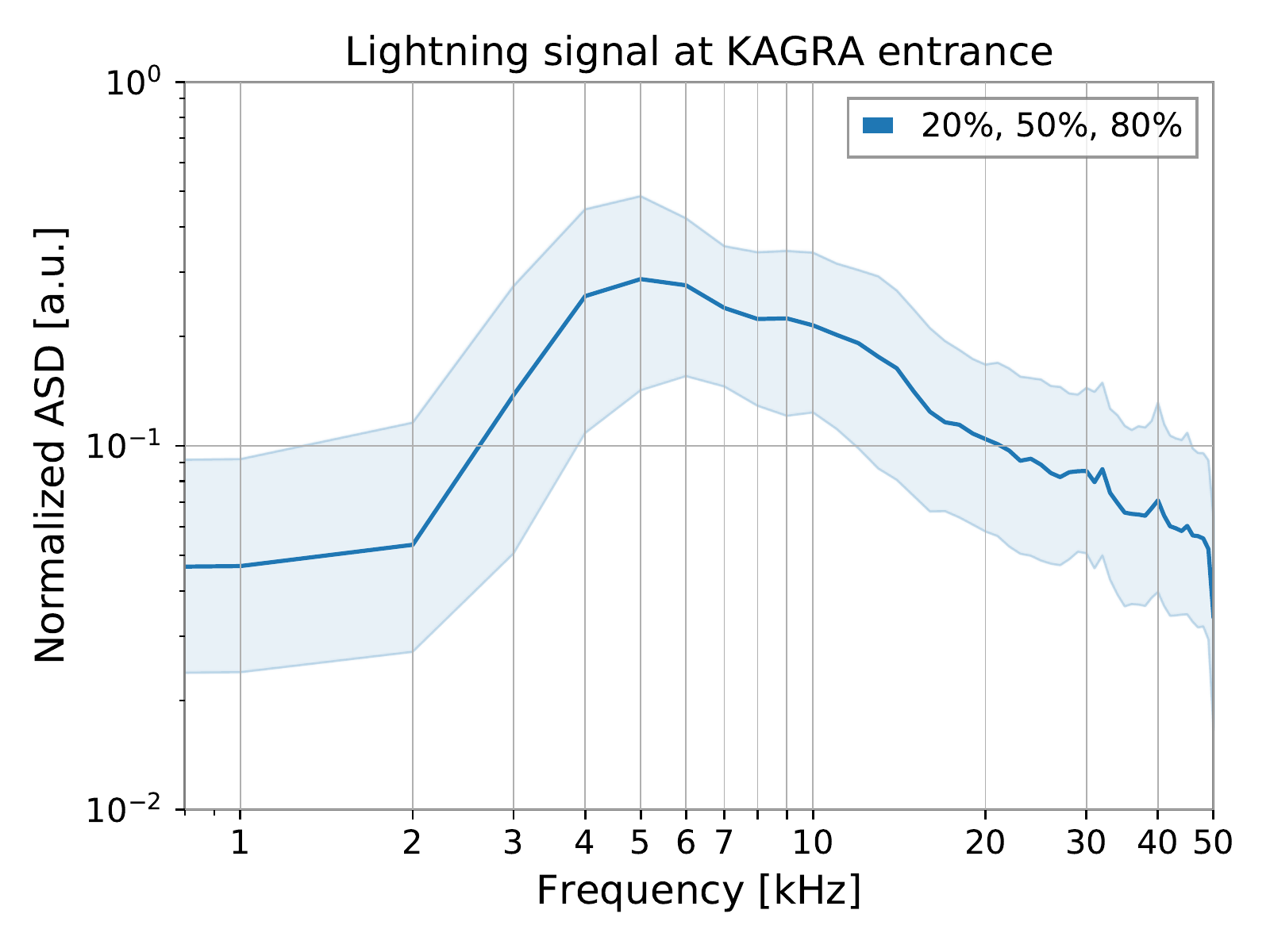}
\caption{\label{fig:No2454} Left: Blitzortung.org detector locations in Japan and No.~2454 at KAGRA entrance.
Right: Percentile of the normalized amplitude spectrum density for the Blitzortung detector (sum of the three axes) No.~2454.}
\end{figure}

Blitzortung.org~\cite{blitzortung,Narita} is a lightning detection and localization network, 
based on time-of-arrival (TOA) and time-of-group-arrival (TOGA) methods for the VLF (3-30~kHz) electromagnetic wave signals.
There are approximately 2,000 detector stations in the world based on the volunteer collaboration. 
A total of 55 detector stations in Japan (Figure~\ref{fig:No2454}, left) and 20 detector stations in the other Asia-Oceania region are managed by the Shonan Institute of Technology~\cite{5656}. 
A typical detection range for a Blitzortung system-blue detector is approximately 4000~km.
Detector station No.~2454 is located in front of the KAGRA tunnel (installed on September 12, 2019). 
The sensing part consists of three ferrite bar antennas and pre-amps (1-50~kHz of the $\rm -3~dB$ range). 
The raw signals of the time series were recorded by a local data logger (GRAPHTEC GL980\textregistered) with a self-trigger, 100~kHz sampling, and 2~s window per events. 
Figure~\ref{fig:No2454} (right) shows the 20, 50, and 80 percentile of the amplitude spectrum density (ASD), normalized by the integral at 1-50~kHz,  
for the lightning events observed from September to December, 2020\footnote{Stable data acquisition started in September 2020}.
It has a peak at approximately 5~kHz, which is consistent with the literature~\cite{Cummins,Bianchi}.

%%%%%%%%%%%%%%%%%%%%%%%%%%%%%%%%%%%%%%%%%%%%%%%%%%%%%%%%%%%%%%%%%%%%%%%%%%%%%%%%%%%%%%%%%%
%\newpage
\section{Lightning signals in the Kamioka underground}
In this section, the lightning signals observed at the KAGRA experimental site in the underground facility are presented, and their characteristics are discussed.

\subsection{Magnetic field}\label{sec:Mag_PSD}
As, it is well known from electromagnetism textbooks, the amplitude of the magnetic field 
through the depth $z$ into a conductor with a resistivity of $\rho$ becomes 
\begin{align}
  B(z) \propto e^{-z/\delta}, \quad    \delta = \sqrt{\frac{\rho}{\pi\mu f}}, \label{eq:slin}
\end{align}
where $\mu$ is the permeability of the conductor, $f$ is the frequency of the magnetic field, 
and $\delta$ is the skin-depth.
From a geophysical survey~\cite{resistivity}, the resistivity of the ground around the Kamioka area 
is measured as $\rho=400~\Omega\mathrm{m}$ and $\mu \simeq \mu_0 = 4\pi\times 10^{-7}$~H/m, respectively.
Then, the skin-depth is calculated as $\delta \simeq 100$~m for $f=1$~kHz. 
The underground depths of the experimental KAGRA site are approximately 200~m and 450~m 
for the CS and X/Y-end stations, respectively. 

\begin{figure}[htbp]
\centering 
\includegraphics[width=15cm,clip]{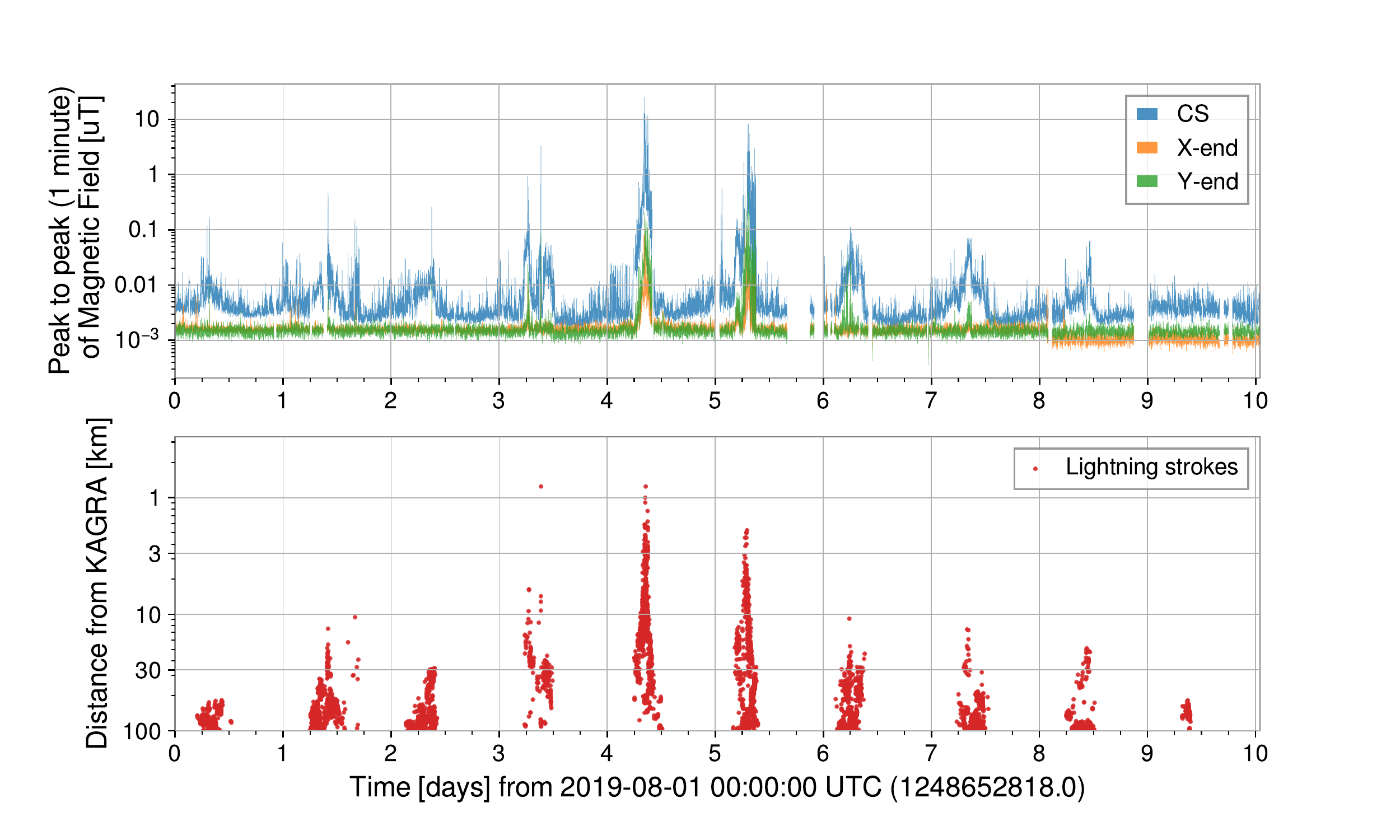}
\caption{\label{fig:mag} Top: The peak-to-peak value of the magnetometers in the KAGRA experimental site for every 1~min.
Bottom: The distance of lightning strokes from KAGRA CS.}
\end{figure}

Figure~\ref{fig:mag} shows an example of the magnetic field (peak-to-peak value every 1~min) 
measured at the KAGRA experimental site and the distance of lightning strokes from KAGRA CS during August 1-10, 2020. 
In this season, we had a large number of lightning occurrences around the Kamioka area almost every evening. 
It can clearly be seen that a large magnetic field was observed when lightning strokes occurred near KAGRA, and this behavior is simply due to the fact that the amplitude of the magnetic wave is decreased in proportion to the path length in the atmosphere. 
The difference in the magnetic strength between sensor positions can be understood from the results of the skin effect, because 
(1) the magnetic strength in the CS is approximately 10~times larger than that in the X- and Y-ends, as expected, and 
(2)  the magnetic strengths in the X- and Y-ends are approximately the same, despite there being an entrance tunnel for the Y-end, but not for the X-end.

Because such behavior is not observed in other types of PEM sensors, such as microphones or accelerometers, 
these peaks are not electrical noise in the data-taking system, but  the real magnetic field variation. 

\begin{figure}[htbp]
\centering 
\includegraphics[width=8.5cm,clip]{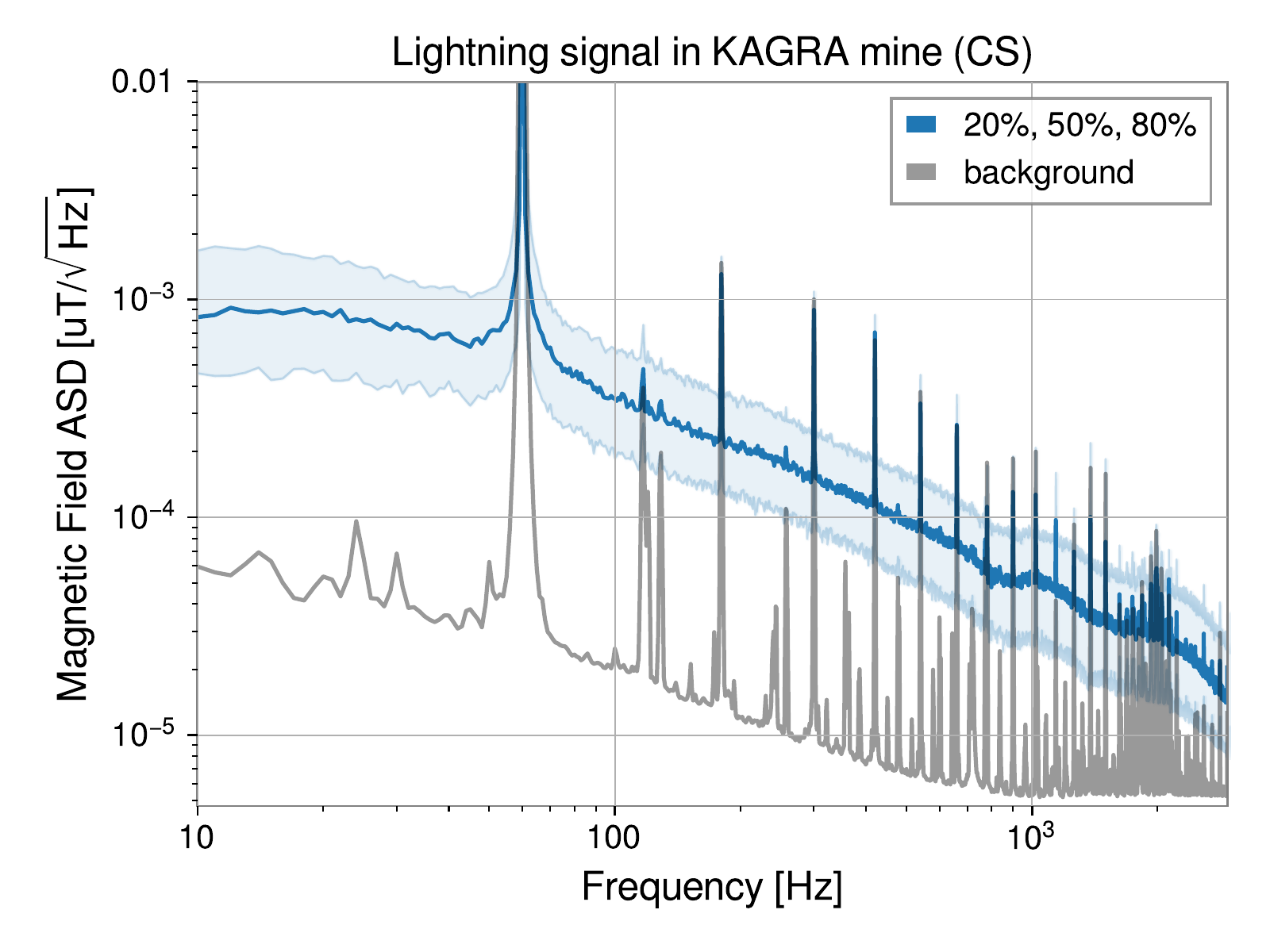}
\caption{\label{fig:Mag13_PSD} ASD of the magnet meter in the KAGRA CS for the lightning events within 30~km (blue) and the background data (gray).}
\end{figure}

The ASDs of the magnet meter located in the KAGRA CS are shown in Figure~\ref{fig:Mag13_PSD}.
The lightning signal (blue) has a monotonous decrease with respect to the frequency (less than 3~kHz), except for the peaks in the background data (gray).
This structure is significantly different from that of the on-ground detector (Figure~\ref{fig:No2454}, right) and is consistent with the skin effect in a qualitative discussion (the magnitude is more reduced for higher frequency), at least in the overlapped frequency of 1-3~kHz.

\subsection{KAGRA main interferometer}\label{sec:3/22event}
\begin{figure}[htbp]
\centering 
\includegraphics[width=11cm,clip]{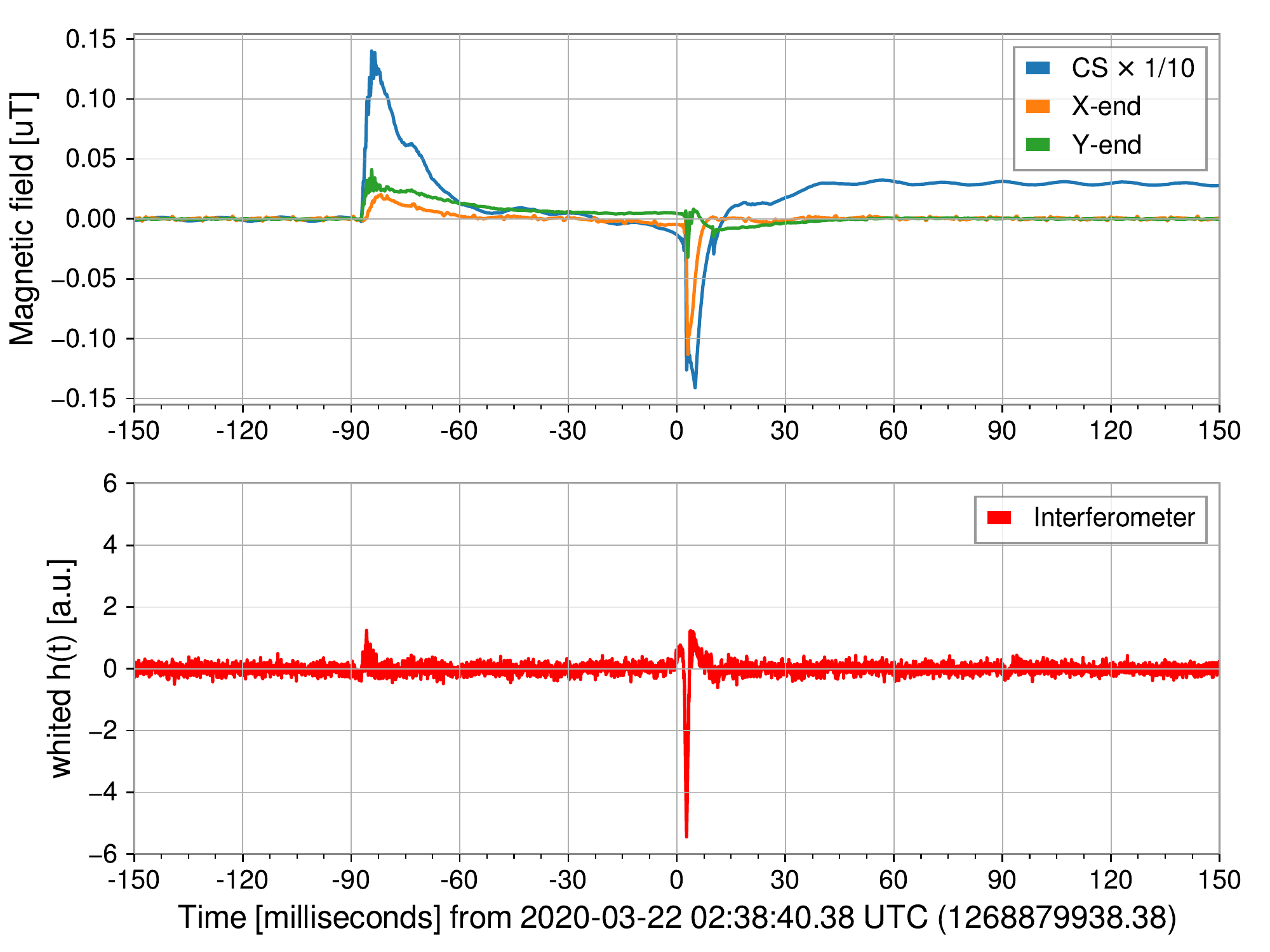}
\caption{\label{fig:TS} Time series of the magnetometers inside of the KAGRA site (top) and the KAGRA main interferometer (whited strain signal) (bottom) for the nearby lightning event. 
For the magnetometers, the DC value of geomagnetism ($\rm\sim 50\mu T$) is subtracted.}
\end{figure}

\begin{figure}[htbp]
\centering 
\includegraphics[width=14cm,clip]{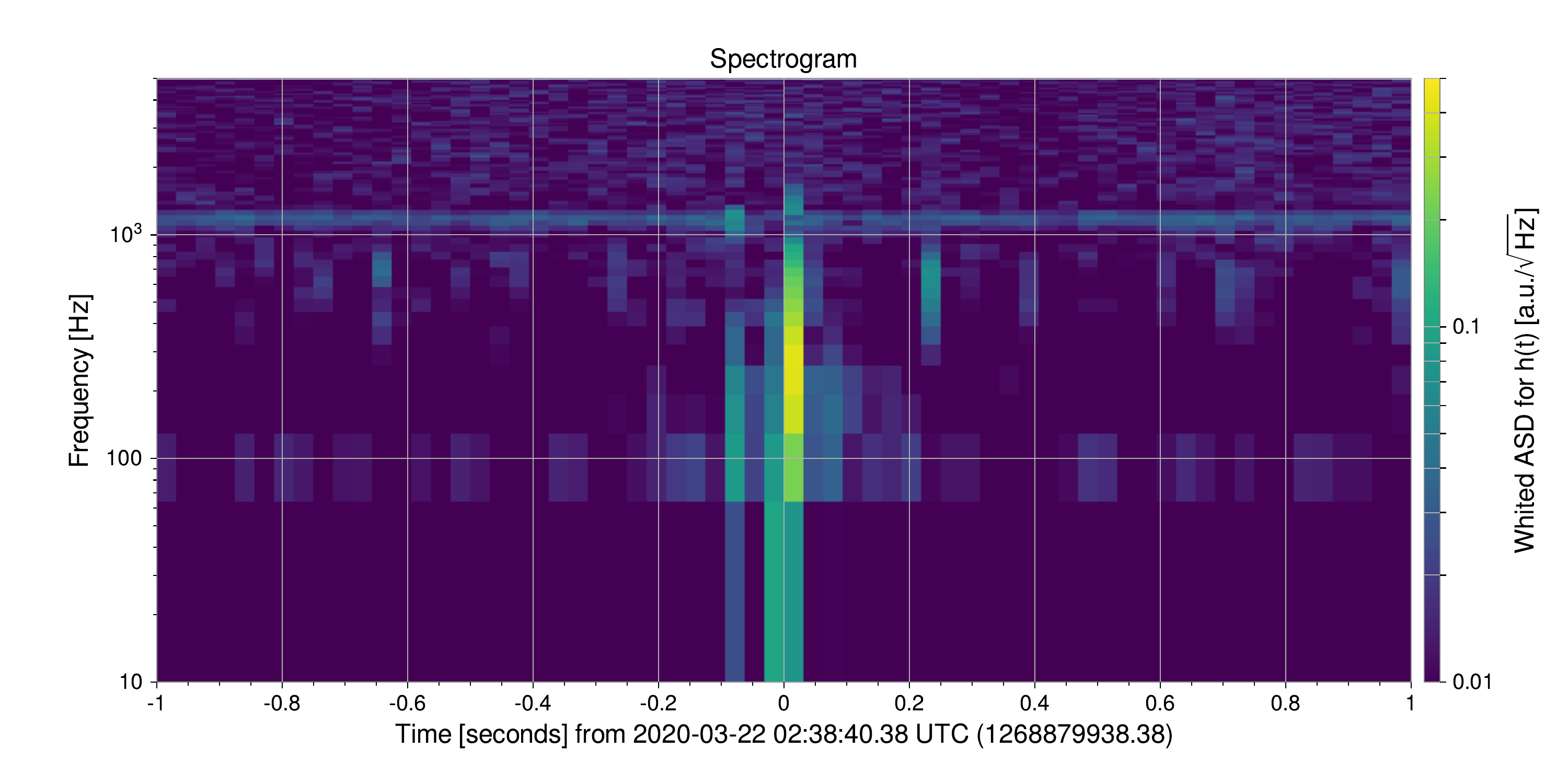}
\caption{\label{fig:SG} The spectrogram of the GW channel (whited) for the lightning event.}
\end{figure}

On 2020-03-22 02:38:40.38 UTC, a lightning stroke occurred extremely close to KAGRA (Figure~\ref{fig:kamioka}), during the locked state of the interferometer. 
This lightning event was detected by both magnetometers and the GW channel (strain sensitivity, $h(t)$) of the main interferometer of KAGRA.
Unfortunately, the local data taking system for the Blitzortung detector was not ready at this time. 
However, no coincident signals were detected by the other auxiliary channels, such as the brink ADC channel (the ADC channel where the data were recorded but nothing was connected), other PEM sensors (seismometers, accelerometers, and microphones), local sensors for the mirror motion, and the beam positions of the interferometer, \textit{etc}. 
Figure~\ref{fig:TS} and Figure~\ref{fig:SG} show their time series and the spectrogram of the GW channel for this event, respectively. 
This is the first evidence that a GW detector constructed in an underground facility is excited by lightning strokes in the atmosphere. 
This means that the lightning will be background events of a burst-GW search, but can be easily identified by the current system of the PEM.


%%%%%%%%%%%%%%%%%%%%%%%%%%%%%%%%%%%%%%%%%%%%%%%%%%%%%%%%%%%%%%%%%%%%%%%%%%%%%%%%%%%%%%%%%%
%\newpage
\section{Possible applications of lightning events}\label{sec:app}
As shown in the previous section, it was pointed out that lightning strokes in the atmosphere are not a desirable phenomena from the viewpoint of GW observation. 
However, we can discuss the effective use of lightning in certain studies.
In this section, we propose two possible applications of lightning events.
\subsection{Simultaneous detection by GW detector and neutrino detector}
It is known that thunderclouds can cause the following interactions with a nucleus in the atmosphere just before a lightning stroke~\cite{e+}:
\begin{align}
& ^{14}N + \gamma \to ^{13}\!N + n,     & \mathrm{(Photonuclear~reaction)}, \label{eq:gamma}\\
& ^{13}N \to ^{13}\!C + e^{+} + \nu_e,  & \mathrm{(\beta^+ ~decay)},         \label{eq:beta}
\end{align}
with $Q=2.22$~MeV and a half-life of $T_{1/2} = 598$~s for $\beta^+$ decay. 
These phenomena have been observed with $\gamma$-ray, but not with neutrinos. 
In the underground facility of Kamioka, the neutrino experiments such as Super-Kamiokande~\cite{SK}, KamLAND~\cite{KamLAND}, and CANDLES~\cite{Candles} are running, and it might be possible for KAGRA to detect lightning events coincident with them. 
One of the most interesting events emitting GW and neutrinos is a supernovae explosion (\textit{e.g., }~\cite{supernovae}) for which the observation of lightning will be a good demonstration. 
Please note that this is a challenging issue because the energy spectrum and time structure are quite different between supernovae and lightning signals. 
In addition, the processes (\ref{eq:gamma}) and (\ref{eq:beta}) are confirmed only for the winter-type lightning (high energy) around the bay area and not for the summer-type lightning in the mountain-area.

%%%%%%%%%%%%%%%%%%%%%%%%%%%%%%%%%%%%%%%%%%%%

\subsection{Estimation of underground water}
If it is correct that the difference in the spectrum of the magnetic field for lightning 
(Figure~\ref{fig:No2454} and Figure~\ref{fig:Mag13_PSD}) is caused by the skin effect, 
the resistivity of the underground can be evaluated by comparing the spectrum of 
broadband magnetometers (\textit{e.g., } DC-10~kHz) located inside and outside the tunnel. 
The ground resistivity $\rho$ corresponds to the amount of water contained in the ground, using Archie's formula~\cite{{Archie}}: 
\begin{align}
    \rho = a \phi^{-m} S^{-n} \rho_w, \label{eq:Archie} 
\end{align}
where $\phi$ is the porosity, $S$ is the brine saturation, $\rho_w$ is the resistivity of water, and $a, m, n$ are constants. 
Underground water information is important for both GW detectors and low-background experiments 
(\textit{e.g.,} dark mater or neutrino-less double-beta decay search), 
because water causes Newtonian gravity noise and the modulation of fast-neutron background~\cite{chisoko}, respectively.

%%%%%%%%%%%%%%%%%%%%%%%%%%%%%%%%%%%%%%%%%%%%%%%%%%%%%%%%%%%%%%%%%%%%%%%%%%%%%%%%%%%%%%%%%%

\section{Conclusion}
\label{sec:conc}
In the KAGRA PEM, a Blitzortung lightning detector and magnetometers were used to monitor the environmental noise for the GW observation. 
This system succeeded in monitoring the magnetic field pulses coming from lightning strokes in the atmosphere, both outside and inside the underground experimental site of KAGRA. 
In addition, it was also confirmed that a lightning-induced event in the GW channel of the KAGRA main interferometer is exhibited, and it can be identified by the current system. 
Lightning events are not only undesirable phenomena but also possible to be applied in some studies. 
\textit{e.g.} demonstration of a supernovae search with neutrino experiments, and estimation for the time dependence of the underground water.

\acknowledgments

This research has made use of data, software, and web tools obtained or developed by the KAGRA Collaboration.
In this study, we were supported by KAGRA collaborators, especially the commissioning members of the interferometer and vibration-isolation systems, administrators of the digital system, and managers of the KAGRA experiment. 
We would like to thank Editage (www.editage.com) for English language editing.

The KAGRA project is funded by the Ministry of Education, Culture, Sports, Science and Technology (MEXT) and 
the Japan Society for the Promotion of Science (JSPS).
Especially this work was founded by
JSPS Grant-in-Aid for JSPS Fellows 19J01299, %% 鷲見学振
JSPS Grant-in-Aid for Scientific Research on Innovative Areas 6105 20H05256, %% 新学術地下宇宙：鷲見
and the Joint Research Program of the Institute for Cosmic Ray Research (ICRR) University of Tokyo 2019-F14, 2020-G12.%, and 2020-G21.


\end{document}